\documentclass[12pt,aps,showpacs,preprintnumbers,superscriptaddress,nofootinbib]{revtex4}
\usepackage{graphicx}

\begin{document}

\def\simge{\hspace*{0.2em}\raisebox{0.5ex}{$>$}
     \hspace{-0.8em}\raisebox{-0.3em}{$\sim$}\hspace*{0.2em}}
\def\simle{\hspace*{0.2em}\raisebox{0.5ex}{$<$}
     \hspace{-0.8em}\raisebox{-0.3em}{$\sim$}\hspace*{0.2em}}
\def\bra#1{{\langle#1\vert}}
\def\ket#1{{\vert#1\rangle}}
\def\coeff#1#2{{\scriptstyle{#1\over #2}}}
\def\undertext#1{{$\underline{\hbox{#1}}$}}
\def\hcal#1{{\hbox{\cal #1}}}
\def\sst#1{{\scriptscriptstyle #1}}
\def\eexp#1{{\hbox{e}^{#1}}}
\def\rbra#1{{\langle #1 \vert\!\vert}}
\def\rket#1{{\vert\!\vert #1\rangle}}
\def\lsim{{ <\atop\sim}}
\def\gsim{{ >\atop\sim}}
\def\nubar{{\bar\nu}}
\def\psibar{{\bar\psi}}
\def\Gmu{{G_\mu}}
\def\alr{{A_\sst{LR}}}
\def\wpv{{W^\sst{PV}}}
\def\evec{{\vec e}}
\def\notq{{\not\! q}}
\def\notk{{\not\! k}}
\def\notp{{\not\! p}}
\def\notpp{{\not\! p'}}
\def\notder{{\not\! \partial}}
\def\notcder{{\not\!\! D}}
\def\notA{{\not\!\! A}}
\def\notv{{\not\!\! v}}
\def\Jem{{J_\mu^{em}}}
\def\Jana{{J_{\mu 5}^{anapole}}}
\def\nue{{\nu_e}}
\def\mn{{m_\sst{N}}}
\def\mns{{m^2_\sst{N}}}
\def\me{{m_e}}
\def\mes{{m^2_e}}
\def\mq{{m_q}}
\def\mqs{{m_q^2}}
\def\mz{{M_\sst{Z}}}
\def\mzs{{M^2_\sst{Z}}}
\def\ubar{{\bar u}}
\def\dbar{{\bar d}}
\def\sbar{{\bar s}}
\def\qbar{{\bar q}}
\def\sstw{{\sin^2\theta_\sst{W}}}
\def\gv{{g_\sst{V}}}
\def\ga{{g_\sst{A}}}
\def\pv{{\vec p}}
\def\pvs{{{\vec p}^{\>2}}}
\def\ppv{{{\vec p}^{\>\prime}}}
\def\ppvs{{{\vec p}^{\>\prime\>2}}}
\def\qv{{\vec q}}
\def\qvs{{{\vec q}^{\>2}}}
\def\xv{{\vec x}}
\def\xpv{{{\vec x}^{\>\prime}}}
\def\yv{{\vec y}}
\def\tauv{{\vec\tau}}
\def\sigv{{\vec\sigma}}
\def\sst#1{{\scriptscriptstyle #1}}
\def\gpnn{{g_{\sst{NN}\pi}}}
\def\grnn{{g_{\sst{NN}\rho}}}
\def\gnnm{{g_\sst{NNM}}}
\def\hnnm{{h_\sst{NNM}}}

\def\xivz{{\xi_\sst{V}^{(0)}}}
\def\xivt{{\xi_\sst{V}^{(3)}}}
\def\xive{{\xi_\sst{V}^{(8)}}}
\def\xiaz{{\xi_\sst{A}^{(0)}}}
\def\xiat{{\xi_\sst{A}^{(3)}}}
\def\xiae{{\xi_\sst{A}^{(8)}}}
\def\xivtez{{\xi_\sst{V}^{T=0}}}
\def\xivteo{{\xi_\sst{V}^{T=1}}}
\def\xiatez{{\xi_\sst{A}^{T=0}}}
\def\xiateo{{\xi_\sst{A}^{T=1}}}
\def\xiva{{\xi_\sst{V,A}}}

\def\rvz{{R_\sst{V}^{(0)}}}
\def\rvt{{R_\sst{V}^{(3)}}}
\def\rve{{R_\sst{V}^{(8)}}}
\def\raz{{R_\sst{A}^{(0)}}}
\def\rat{{R_\sst{A}^{(3)}}}
\def\rae{{R_\sst{A}^{(8)}}}
\def\rvtez{{R_\sst{V}^{T=0}}}
\def\rvteo{{R_\sst{V}^{T=1}}}
\def\ratez{{R_\sst{A}^{T=0}}}
\def\rateo{{R_\sst{A}^{T=1}}}

\def\mro{{m_\rho}}
\def\mks{{m_\sst{K}^2}}
\def\mpi{{m_\pi}}
\def\mpis{{m_\pi^2}}
\def\mom{{m_\omega}}
\def\mphi{{m_\phi}}
\def\Qhat{{\hat Q}}

\def\FOS{{F_1^{(s)}}}
\def\FTS{{F_2^{(s)}}}
\def\GAS{{G_\sst{A}^{(s)}}}
\def\GES{{G_\sst{E}^{(s)}}}
\def\GMS{{G_\sst{M}^{(s)}}}
\def\GATEZ{{G_\sst{A}^{\sst{T}=0}}}
\def\GATEO{{G_\sst{A}^{\sst{T}=1}}}
\def\mdax{{M_\sst{A}}}
\def\mustr{{\mu_s}}
\def\rsstr{{r^2_s}}
\def\rhostr{{\rho_s}}
\def\GEG{{G_\sst{E}^\gamma}}
\def\GEZ{{G_\sst{E}^\sst{Z}}}
\def\GMG{{G_\sst{M}^\gamma}}
\def\GMZ{{G_\sst{M}^\sst{Z}}}
\def\GEn{{G_\sst{E}^n}}
\def\GEp{{G_\sst{E}^p}}
\def\GMn{{G_\sst{M}^n}}
\def\GMp{{G_\sst{M}^p}}
\def\GAp{{G_\sst{A}^p}}
\def\GAn{{G_\sst{A}^n}}
\def\GA{{G_\sst{A}}}
\def\GETEZ{{G_\sst{E}^{\sst{T}=0}}}
\def\GETEO{{G_\sst{E}^{\sst{T}=1}}}
\def\GMTEZ{{G_\sst{M}^{\sst{T}=0}}}
\def\GMTEO{{G_\sst{M}^{\sst{T}=1}}}
\def\lamd{{\lambda_\sst{D}^\sst{V}}}
\def\lamn{{\lambda_n}}
\def\lams{{\lambda_\sst{E}^{(s)}}}
\def\bvz{{\beta_\sst{V}^0}}
\def\bvo{{\beta_\sst{V}^1}}
\def\Gdip{{G_\sst{D}^\sst{V}}}
\def\GdipA{{G_\sst{D}^\sst{A}}}
\def\fks{{F_\sst{K}^{(s)}}}
\def\FIS{{F_i^{(s)}}}
\def\fpi{{F_\pi}}
\def\fk{{F_\sst{K}}}

\def\RAp{{R_\sst{A}^p}}
\def\RAn{{R_\sst{A}^n}}
\def\RVp{{R_\sst{V}^p}}
\def\RVn{{R_\sst{V}^n}}
\def\rva{{R_\sst{V,A}}}
\def\xbb{{x_B}}

\def\PR#1{{{\em   Phys. Rev.} {\bf #1} }}
\def\PRC#1{{{\em   Phys. Rev.} {\bf C#1} }}
\def\PRD#1{{{\em   Phys. Rev.} {\bf D#1} }}
\def\PRL#1{{{\em   Phys. Rev. Lett.} {\bf #1} }}
\def\NPA#1{{{\em   Nucl. Phys.} {\bf A#1} }}
\def\NPB#1{{{\em   Nucl. Phys.} {\bf B#1} }}
\def\AoP#1{{{\em   Ann. of Phys.} {\bf #1} }}
\def\PRp#1{{{\em   Phys. Reports} {\bf #1} }}
\def\PLB#1{{{\em   Phys. Lett.} {\bf B#1} }}
\def\ZPA#1{{{\em   Z. f\"ur Phys.} {\bf A#1} }}
\def\ZPC#1{{{\em   Z. f\"ur Phys.} {\bf C#1} }}
\def\etal{{{\em   et al.}}}

\def\delalr{{{delta\alr\over\alr}}}
\def\pbar{{\bar{p}}}
\def\lamchi{{\Lambda_\chi}}

\title{Parity-Violating Electron Scattering: How Strange a Future
\footnote{
Talk given at PAVI02, workshop on parity violation in electron scattering,
Mainz,
Germany, June 2002. }}

\author{M.J. Ramsey-Musolf}
\affiliation{
California Institute of Technology,
Pasadena, CA 91125\ USA}
\affiliation{
Department of Physics, University of Connecticut,
Storrs, CT 06269\ USA}

\begin{abstract}
I discuss several physics issues that can be addressed through
the present
and future
program of parity-violating electron scattering measurements. In
particular, I focus
on strange quark form factors, hadronic effects in electroweak radiative
corrections,
and physics beyond the Standard Model.
\end{abstract}

%\pacs{14.20.Dh, 11.55.-m, 11.55.Fv}

%\date

\maketitle

\section{Introduction}

Parity-violating electron scattering (PVES) has  become an
established component of the
nuclear physics research program at a variety of accelerators. Prior to the
late 1990's, the
field was considered somewhat esoteric, with the completion of roughly one
experiment per
decade: the deep inelastic $eD$ experiment at SLAC in the
1970's\cite{prescott}; the
quasielastic beryllium
experiment at Mainz in the 1980's\cite{heil}; and the elastic carbon
measurement
culminating in a
publication in 1990\cite{souder}. In the last few years, the rate at which
new results
are being reported
has accelerated dramatically compared that of the past 30 years, and in the
next several
years, we anticipate even more results from the SAMPLE Collaboration at
MIT-Bates\cite{sample1}, the
Happex\cite{aniol}, G0\cite{beck}, PAVEX\cite{michaels}, and
Q-Weak\cite{carlini}
Collaborations at Jefferson Lab, the A4 experiment at Mainz\cite{maas},
and the E158 experiment at SLAC\cite{hughes}. Looking further down the
road, the
prospective up-grade of
CEBAF may open up additional possibilities for PVES studies beyond those
currently on the
books. Suffice it to say, PVES has come into its own as an important tool
for the study of a
broad range of questions in nuclear and particle physics, moving far beyond
its earlier
incarnation as an almost exotic area of physics (for a review, see Ref.
\cite{musolf94}).

At the end of the day, what physics do we hope to have accomplished with
this tool, and what
important questions do we hope the field will have answered? In this talk,
I would like to
give my view. At the same time, I will discuss some recent developments
that have opened up
new directions for study of both hadron structure and particle physics. Of
course, this
theorist's perspective will be weighted in favor of new theory, but I want
to emphasize that
these theoretical developments would have only academic interest were it
not for the
substantial progress on the experimental side. In fact, one of the
attractive features of the
PVES community is the close interplay between theory and experiment. The proper
interpretation of the very tiny asymmetries measured with PVES requires
careful theoretical
delineation of various contributions. Conversely, the judicious choice of
target and
kinematics, together with hard-nosed experimental study of the systematic
effects that could
generate \lq\lq false" asymmetries, provide theorists with a clean and
unambiguous probe of nucleon
structure and the weak interaction. This kind of syngergy between theory
and experiment is a
hallmark of our field.

\section{Strange quarks}

The present burst of activity in the field has been motivated by a desire
to probe the
strange quark content of the nucleon. Understanding strange quarks is
important from a number
of perspectives:

\medskip

\noindent {\em Flavor decomposition of nucleon properties}. How much do the
different
flavors of light quarks contribute to the low-energy properties of the
nucleon? For many
years, the standard lore was that up- and down-quark degrees of freedom
were sufficient to
account for the most familiar aspects of the nucleon, such as its mass,
spin, and magnetic
moment. Analyses of the $\pi$N $\sigma$-term suggested that light-quarks
were not, in fact,
the whole story, and that possibly 20-30\% of $m_N$ came from $s{\bar s}$
pairs\cite{gasser}. Later,
polarized deep inelastic scattering (DIS) measurements implied that only
30\% of the nucleon
spin came from quark spin, with the remainder being supplied by quark
orbital angular momentum
and by gluons\cite{filippone}. Part of the picture associated with the
\lq\lq spin-crisis" was that
the fraction of nucleon spin coming from $s{\bar s}$ pairs was about 10\%,
and that these pairs
were polarized opposite to the direction of the total spin. In the quark
model framework,
this idea is a complete mystery. As a result, one would like to know what
other nucleon
properties are substantially affected by strangeness. Hence, the current
program of
measurements aimed at the strange quark contributions to nucleon
electromagnetic properties.

The PVES strange quark program offers one advantage over other flavor
decomposition studies,
namely, the theoretically clean character of interpretation. Obtaining the
$s{\bar s}$
contribution to the nucleon mass requires extrapolation of $\pi$N
scattering amplitudes into
the unphysical regime using a combination of dispersion relations and
chiral perturbation
theory (ChPT). The value for the $\bra{N}{\bar s}s\ket{N}$ -- relative to
$\bra{N} {\bar u}u
+{\bar d}d\ket{N}$ -- has varied somewhat over the years, in
part due to the ambiguities associated with this extrapolation.

The situation regarding the
strange quark contribution to the nucleon spin, $\Delta s$, is even more
ambiguous. The
extraction of $\Delta s$ from polarized DIS measurements relies on the
assumption of flavor SU(3)
symmetry. The relevant matrix element is not measured directly in the
experiments, and
information on neutron $\beta$-decay and hyperon semileptonic decays must
be incorporated
into the analysis using SU(3) symmetry.  For some time, people have worried
about the
uncertainties associated with SU(3)-breaking. Recently, my collaborators
and I studied these
effects in ChPT and found them to be potentially serious\cite{zhu1,zhu2}.
Specifically, one has
\begin{eqnarray}
\label{eq:deltas}
\Delta s & = & \frac{3}{2} \left[ \Gamma_p + \Gamma_n\right]
-\frac{5\sqrt{3}}{6} g_A^8 \\
\nonumber
& = & 0.14 - [0.12+0.25 +0.10] \ \ \ ,
\end{eqnarray}
where the $\Gamma_N$ are the integrals of $g_1^N(x)$ taken from experiment
and $g_A^8$ is
the eighth component of the octet of axial current matrix elements. A value
for the latter
can only be obtained from experimentally measured axial current matrix
elements by invoking
SU(3) symmetry. The chiral expansion of this SU(3) relationship is shown in
the square
brackets in the second line of Eq. (\ref{eq:deltas}) (the first term is the
experimental
value for the sum of $g_1^N$ integrals). The first, second, and third terms
represent the ${\cal
O}(p^0)$, ${\cal O}(p^2)$, and ${\cal O}(p^3)$ contributions, respectively.
It is a manifestly
non-converging series. Based on the formula, we have no way of estimating
the theoretical
SU(3)-breaking uncertainty in $\Delta s$, nor can we really say what the
value of $\Delta s$
is.

The advantage of PVES is that the interpretation of the results in terms of
strange quarks
does not suffer from these kinds of ambiguities. Once we have sufficiently
precise
measurements of the proton and neutron electromagnetic and weak neutral
current form factors, we know
the relative contributions of $u$, $d$, and $s$-quarks to the nucleon
vector current matrix
elements. The relative importance of the strange quark contribution may be
either large or
small, but either way, the answer will be clear. We will know the flavor
decomposition of the
nucleon vector current matrix elements\footnote{Kaplan and Manohar showed
some time ago that
the contributions from heavier quarks scale as $(\Lambda_{\rm QCD}/M_Q)^4$
and are, thus,
negligible\cite{kaplan}.}.

\medskip

\noindent{\em Sea quarks and the quark model}. Why does the constituent
quark model work so
well in describing low-energy nucleon properties? In light of what we know
about the presence
of sea quarks and gluons in the nucleon, the success of the constituent
quark description of
the nucleon remains a puzzle. A variety of solutions have been proposed. It
may be that sea
quarks are simply \lq\lq inert" when viewed over long-distance scales. A
more interesting
possibility is that the sea quarks are not inert but that their effects are
hidden in
the effective parameters of the quark model, such as the constituent quark
mass and the
\lq\lq string tension" of the ${\bar q} q$ potential\footnote{I am indebted
to Nathan Isgur
for my understanding of this idea.}. It may also be that nature is more
devious than either
of these other possibilities suggest and has conspired to conceal the
effects of sea quark
dynamics through a series of cancellations. The latter view seems to emerge
from dispserion
relation analyses of isoscalar electromagnetic and strange quark form
factors, as I discuss
below. In any case, the beauty of the PVES measurements is that they probe
the $s{\bar s}$ sea
over distance scales relevant to the quark model description. If either the
effective quark
model parameters or nature itself has hidden sea quark effects, the
measurements will lift
the veil of secrecy and allow us to see what the sea is up to at low-energies.

\medskip

\begin{figure}
\begin{center}
\resizebox{5. cm}{!}{\includegraphics*[200,400][420,730]{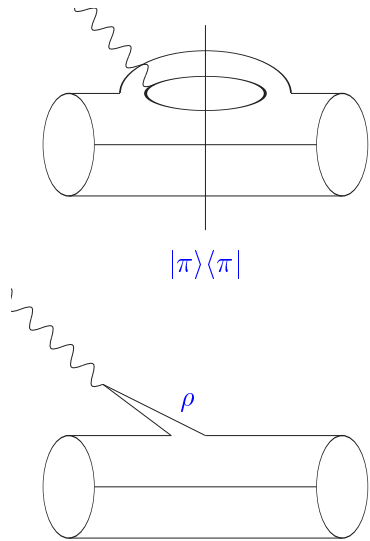}}
\caption{\label{Fig1}Disconnected quark loop and Z-graph contributions to
the nucleon isovector
electromagnetic form factors.}
\end{center}
\end{figure}

\noindent{\em Z-Graphs vs. Loops}. How do we relate the quark model
description of the nucleon
with one built out of hadronic degrees of freedom? In the latter framework,
for example, the
low-momentum, isovector electromagnetic properties of the nucleon involve a
combination of
the pion cloud and $\rho$-meson resonance. At the most naive level, one may
think of the
$\rho$-meson resonance as a \lq\lq Z-graph" effect and the pion cloud as
involving
disconnected quark loops (see Fig. \ref{Fig1}). In the $t$-channel, the
$\rho-\gamma$ coupling
involves $q{\bar q}$ pair creation, and in order to connect to a nucleon,
it must also involve
the negative frequency part of the constituent quark propagator (the
Z-graph). In the same
channel, the pion cloud contains a $\pi\pi$ intermediate state. At the
quark level, the
non-resonant part of this contribution involves the presence of two $q{\bar
q}$ pairs. With
only three constituent quarks in the nucleon, this effect looks like a
disconnected quark
loop -- a sea quark effect. On the other hand, in the non-relativistic
quark model (NRQM),
neither Z-graphs or loops are present, and yet one can still reproduce the
low-momentum
isovector properties of the nucleon. How can this be?

The strange quark contributions to the vector current form factors
necessarily involve
disconnected quark loops, since there are no valence strange quarks in the
constituent quark
model nucleon wavefunction. In a hadronic langauge, these loops look, for
example, like $K{\bar K}$
pairs. Dispersion relations also suggest that this kaon cloud is actually
dominated by the $\phi$
resonance -- a Z-graph effect that would imply more sizeable strangeness
form factors than
one gets from loops only. If this picture is confirmed by experiment, then
it will force us
to think more carefully about how to relate the quark and hadronic
descriptions of the
nucleon.

\medskip

\noindent{\em Perturbation theory may not apply}. The mass of the strange
quark forces us
into a theoretical no man's land where we cannot easily apply our usual
perturbative
techniques for making predictions. The strange quark is not heavy enough to
use heavy quark
effective theory; since $\Lambda_{\rm QCD}/m_s\sim 1$, we have no small
expansion parameter
as we do for heavier quarks. At the same time, the strange quark may not be
light enoughto
make SU(3) chiral perturbation theory applicable; $m_K/\Lambda_\chi\sim
1/2$ -- not a
particularly small expansion parameter either. Thus, from the theoretical
side, the study of
strange quarks presents a particularly challenging situation.

\medskip

\noindent{\em Symmetry is impotent}. Because the strange quark vector
current contains a
component that is a flavor SU(3) singlet, we cannot use SU(3) symmetry to
relate the
strangeness form factors to any other presently measured quantities.
Specifically\cite{ito}
\begin{equation}
{\bar s}\gamma_\mu s = J_\mu^B -2 J_\mu^{I=0}\ \ \ ,
\end{equation}
where $J_\mu^B$ is the baryon number, or SU(3) singlet current, and
$J_\mu^{I=0}$ is
the isoscalar electromagnetic (EM) current. We know quite a bit about the
latter, but
the PVES measurements will, in fact, give us our first information on the
form factors
associated with $J_\mu^B$. This means that, even if the SU(3) chiral
expansion were
well-behaved, there would appear counterterms or low-energy constants
(LEC's) that we
cannot obtain from other experiments. Consequently, we cannot in general
use ChPT to predict
$\GES$ and $\GMS$, the strangeness electric and magnetic form factors,
respectively.

\medskip

Clearly, the study of these form factors will lead to new insights into
nucleon structure,
and PVES provides us with a theoretically clean probe of this physics. The
theoretical
challenge is to translate the results of PVES measurements into a clear
understanding of sea
quark dynamics at distance scales for which traditional methods of treating
nucleon
structure run into trouble.

\subsection{Nonperturbative methods: recent developments and open questions}

Generally speaking, theorists have taken one of three approaches to the
study of $\GES$ and
$\GMS$: hadron models, dispersion relations, and lattice QCD. Of these, the
latter two
represent first principles methods at the hadronic and microscopic levels,
respectively.
Models have their place, but  they necessarily involve assumptions and
approximations
that are not always clearly identified or controlled. For this reason, we
have witnessed a
substantial number of model predictions for $\GES$ and $\GMS$ and a
correspondingly wide
range of values. Should experiment agree with any subset of these
calculations, it will be
hard to gauge the significance of this agreement. For this reason, my bias
favors the first
principles treatments, even if they cannot yet be implemented in a
completely successful way.

Although -- as I emphasized above -- ChPT cannot provide predictions for
$\GES$ and $\GMS$,
it nevertheless provides a convenient framework in which to discuss the
physics coming out
of other non-perturbative methods. For concreteness, I will focus on $\GMS$
at small
$q^2$. In this kinematic domain, one has an expansion
\begin{equation}
\GMS(q^2) = \mu_s + \frac{1}{6}<r^2_s>_M q^2 +\cdots\ \ \ .
\end{equation}
The strangeness magnetic moment $\mu_s$ is given by\cite{ito,meissner,hammer1}
\begin{equation}
\mu_s = \left({2 M_N\over\lamchi}\right)b_s +\cdots
\end{equation}
where $b_s$ is an {\em a priori} unknown low-energy constant that governs
the lowest
order [${\cal O}(p^2)$] contribution and the $+\cdots$ denote higher order
contributions
arising from loops and subleading operators. As noted earlier, one cannot
use existing
measurements to predict $\mu_s$ because $b_s$ contains an SU(3) singlet
component.

In contrast, the lowest order [${\cal O}(p^3)$] contribution to the
magnetic radius arises
entirely from chiral loops\cite{meissner,hammer1}:
\begin{equation}
<r^2_s>_M = -{\pi M_N\over 3
m_K\Lambda_\chi^2}\left(5D^2-6DF+9F^2\right)+\cdots
\end{equation}
where $D$ and $F$ are the usual SU(3) reduced matrix elements and the
$+\cdots$ denote higher
order contributions. Since $D$ and $F$ are known from other experiments, to
${\cal
O}(p^3)$ we have a
parameter-free prediction for the strange magnetic radius. This result has
been used to
extrapolate the SAMPLE determination of $\GMS$ at $Q^2=-q^2=0.1$
(GeV$/c)^2$ to the photon
point in order to obtain results for $\mu_s$\cite{sample1}.

Given the size of the expansion parameter $m_K/\lamchi$, however, one might
worry about
the possible importance of higher-order contributions. To that end, the
${\cal O}(p^4)$
contributions to $<r^2_s>_M$ were recently computed in Ref. \cite{hammer1}.
The loop effects arising
at this order nearly cancel those appearing at ${\cal O}(p^3)$, and one
also encounters a
new LEC. The resulting expression is
\begin{equation}
<r^2_s>_M = -[0.04+0.3 b_s^r] \ \ {\rm fm}^2\ \ \ ,
\end{equation}
where the first term in the square brackets gives the loop contribution and
$b_s^r$ is the
${\cal O}(p^4)$ magnetic radius LEC. On general grounds, one would expect
the magnitude of
$b_s^r$ to be of order unity with undetermined sign. In principle, then,
its contribution
could be an order of magnitude more important than the loop contribution.
Until we are able
to determine its value, we really do not know the either the magnitude or
the sign of the
slope of $\GMS$ at the origin. As a result, the determination of $\GMS$ at
a single
kinematic point is not sufficient to determine $\mu_s$.

This new development has a couple of implications. First, in order to learn
how much the
strange quarks contribution to the nucleon magnetic moment, we will require
an experimental
determination of the $q^2$-dependence of $\GMS$ at low-$q^2$. Accomplishing
this determination
will require completion of the the backward angle, low-$q^2$ program as
envisioned for the
G0 experiment. Second, it means that understanding the non-perturbative
dynamics of the
strange sea -- as parameterized by $b_s$ and $b_s^r$ -- will be key to
achieving a
theoretical understanding of the strange magnetism.

The strangeness electric form factor has a similar chiral expansion. Since
the nucleon has no net
strangeness, the leading $q^2$-dependence is governed by the strangeness
electric radius,
$<r^2_s>_E$. In ChPT, this term arises at ${\cal O}(p^3)$, and both loops
and an electric
radius LEC -- $c_s^r$ -- appear at this order\cite{ito}:
\begin{equation}
\langle r^2_s\rangle_E \approx [-0.15+0.17 c_s^r]\ \ {\rm fm}^2\ \ \  .
\end{equation}
As in the case of the strangeness magnetic
moment and radius, we require input from non-perturbative methods in order
to obtain predictions for
$<r^2_s>_E$.

\medskip

\noindent{\em Lattice QCD}. In principle, lattice QCD computations provide
a means of obtaining
values for the constants $b_s$, $b_s^r$, and $c_s^r$. The two most recent
lattice computations
provide conflicting results for these constants. Both computations were
carried out in the
quenched approximation, so that $s{\bar s}$ pairs appear only via the
operator insertion. Even
under this quenched approximation, computation of such \lq\lq disconnected
loop" contributions is
difficult. For connected insertions, wherein the operator is inserted on a
line originating from
the hadronic source, one must compute quark propagators from a fixed point
in spacetime to all
possible points on the lattice. In contrast, disconnected insertions
require computation of quark
propagators from all points on the lattice  to all other points on the
lattice. Carrying out such
a computation is inordinately expensive. To get around this barrier, one
employs \lq\lq noise"
methods, in which one computes quark propagators from a random selection of
initial points to
a random set of final spacetime points\cite{noise}. For a sufficiently
large number of noise vectors,
operator matrix elements computed using these quark propagators will
approach the value one would
obtain by computing propagators $S(x,x')$ for all $x$ and $x'$.

The two recent lattice computations adopted different philosophies for
treating the noise method.
The Kentucky-Adelaide group employed a relatively small number of gauge
field configurations but a
large number of noise vectors\cite{liu1}. They obtained statistically
signficant signals for $\GES$
and $\GMS$ at a variety of $q^2$ values, and from these results, one may
infer values for the
strangeness LEC's. In contrast, the computation of Ref. \cite{lewis} used
on the order of 1000 gauge
field configurations but a relatively small number of noises. In this
calculation, no significant
signal was observed for the strangeness vector current matrix elements, but
a non-zero value for
the scalar density $\bra{N} {\bar s} s\ket{N}$ was obtained. Resolution of
this disagreement will
require signficant future work, and the two groups are pursuing new,
refined calculations of the
strangeness form factors. Until a consensus emerges, it is difficult to
draw any strong
conclusions from lattice computations. For purposes of illustration,
however, I will use the
results from Ref. \cite{liu1} below.

Achieving firm lattice QCD predictions for $\bra{N}{\bar s}\gamma_\mu
s\ket{N}$ will ultimately
require resolution of a number of issues. Clearly, one would like to have
in hand an unquenched
calculation. In addition, one would like to use sufficiently small values
for the light quark
masses that one can extrapolate to the physical values using ChPT. In this
respect, use of
chiral fermions -- either domain wall or overlap -- is likely to be
critical. Recent work by the
Kentucky group suggests that one must use very light quarks in order for
the chiral extrapolation
to be accurate, and in this respect, overlap fermions may provide some
advantage\cite{liu2}.

\medskip

\noindent{\em Dispersion Relations}. Athough the use of dispersion
relations is not a microscopic
method, it nevertheless provides a first principles approach to computing
form factors in a
hadronic basis. The inputs required include the analyticity properties of
form factors, causality,
unitarity, and experimental scattering amplitudes. To be concrete, the
dispersion relation for the
strangeness magnetic and electric radii is
\begin{equation}
\label{eq:disprel}
<r^2_s>_{M,E} = {6\over\pi} \ \int_{9 m_\pi^2}^\infty {{\rm Im}
G_{M,E}^{(s)}(t)\over t^2}\ dt
\ \ \ .
\end{equation}
The spectral function ${\rm Im} G_{M,E}^{(s)}(t)$ can be related to
experimental scattering
amplitudes by drawing upon some basic ideas in field theory. Symbolically,
one has
\begin{equation}
\label{eq:spectral}
{\rm Im} G_{M,E}^{(s)}(t)\sim {\cal P}_{M,E} \sum_n <N{\bar N}| n> <n|
{\bar s}\gamma_\mu s |0>\ \
\ ,
\end{equation}
where ${\cal P}_{M,E}$ are magnetic and electric projection operators and
the states $\ket{n}$ are
all states carrying the quantum numbers of ${\bar s}\gamma_\mu s $. The
lighest such state is the
$3\pi$ state. Although pions themselves have no valence strange quarks,
three pions can resonate
into an $\omega$ and, by virtue of $\omega$-$\phi$ mixing, can give
non-zero contributions to
the strangeness vector current matrix elements. We have no indication that
the $5\pi$ and
$7\pi$ states resonate into an ${\bar s}s$ vector meson, so they are
usually omitted from the
analysis\footnote{This approximation could entail introduction of some
error, and it should be
scrutinized further.}.

In order of mass, the next state -- and the most intuitive contributor --
is the $K^+K^-$ state.
The ChPT expressions for loop contributions to $\mu_s$ and $<r^2_s>_{M,E}$
are derived by
computing the $K^+K^-$ contribution at one-loop order. However, the work of
Ref. \cite{hammer2}
implies that the one-loop approximation is physically unrealistic. In order
to satisfy the
requirement of S-matrix unitarity, one must sum up kaon rescattering
corrections to {\em all orders}.
After doing so, one finds that the kaon cloud contribution to the spectral
functions is dominated by
the
$\phi(1020)$ resonance. Moreover, the presence of this resonance
signficantly enhances the kaon
cloud contribution to the form factors.

An interesting picture of form factor dynamics then emerges. To illustrate,
consider the
isoscalar and strangeness magnetic moments. In the case of $\mu^{I=0}$, the
$3\pi$ and
$K^+K^-$ contributes have nearly equal magnitude but opposite sign, and the
resulting
cancellation leads to the very small value for the isoscalar moment. For
$\mu_s$, in contrast,
the $3\pi$ contribution gets suppressed by the $\omega$-$\phi$ mixing angle
($\epsilon\sim 0.05$),
while the kaon cloud contribution gets enhanced roughly by the inverse of
the strange quark's
electric charge, or $-3$. The resulting value for $\mu_s$ is roughly $-0.3$
-- sizeable enough
to be seen by experiment. In short, nature conspires to hide the effects of
resonating $s{\bar s}$
pairs in $\mu^{I=0}$ through a cancellation against resonating light quark
pairs. Far from being
inert, the $s{\bar s}$ sea is quite active, though hidden from view in
purely electromagnetic (EM)
processes. Measuring the strangeness form factors directly should allow one
to uncover the
conspiracy.

There is one possible loop hole in this picture. So far, we have only been
able to include states
up through the $K^+K^-$ state in a rigorous way. The limitation is the
absence of strong
interaction scattering data involving higher-mass states. If the isovector
EM form factors are any
guide, then the low-mass states ought to  suffice to describe the leading
strangeness moments.
Indeed, the non-resonant $\pi\pi$ continuum and $\rho$-meson resonance
give nearly all
of the low-$q^2$ behavior of the isovector form factors. For the
strangeness form factors,
however, there is no guarantee that inclusion of the low-mass region is
enough. Considerations of
the large $q^2$-behavior of these form factors along with simple quark
counting arguments suggests
that higher-mass states ({\em e.g.}, $KK\pi$, {\em etc}.) may be required.
Unfortunately, our
ability to carry out a dispersion theory treatment of the higher-mass
region is presently limited
by the dearth of data.

On the other hand, one could ultimately intepret the results of the $\GES$
and $\GMS$ measurements in terms of the relative importance of this
higher-mass region. If, for
example, the value of $\mu_s$ differs substantially from $-0.3$ as
suggested by the SAMPLE results,
then there would have to be important higher-mass effects in order to
overcome the fairly
sizeable low-mass contribution. Moreover, these higher-mass effects would
have to be relatively
strong for the strangeness current but relatively small for the isoscalar
EM current. Such a
situation would suggest that nature is an even more devious conspirator
than we might otherwise
think.

\medskip

\noindent{\em Hadronic models}. While I cannot review the full breadth of
hadron model predictions
for $G_{M,E}^{(s)}$, a few comments are warranted. From the standpoint of
the dispersion theory
treatment described above, many hadronic models are based on the idea of
kaon cloud dominance.
Intuitively, this idea is appealing. It says the strange sea gets polarized
because the nucleon
fluctuates into a kaon-hyperon pair. Many calculations -- including some I
carried out in the past --
relied on one-loop computations to estimate this effect. The dispersion
relation analysis,
however, implies that the picture is physically unrealistic. Inclusion of
rescattering effects,
encoded by the LEC's $b_s$, $b_s^r$, {\em etc.}, is essential. Moreover,
$s{\bar s}$ pair creation
likely plays a more important role than the spatial separation of an $s$
and ${\bar s}$. Finally,
one may encounter cancellations between various contributions. In this
respect, the quark model
calculation of Geiger and Isgur is suggestive\cite{geiger}. In that work, a
sum over a tower of two
hadron intermediate states was considered, with the quark model providing
the relevant couplings. As
one proceeds up the tower, strong cancellations occur between various
contributions, leading to small
strangeness form factors in the end. By necessity, however, the calculation
was carried out to
one-loop level only. Whether this pattern of cancellations would persist
after rescattering and
resonance effects were included is hard to guess.

\medskip

\noindent{\em Some Numbers}. While the task of computing strange quark form
factors has a long way
to go, it is nonetheless useful to discuss some of the numbers coming from
the calculations. In
Table I, I give some of the lattice QCD and dispersion relation values for
the LEC's $b_s$,
$b_s^r$, and $c_s^r$. These numbers should be taken with a grain of salt.
As noted above, the two
most recent lattice QCD computations differ substantially, with one seeing
no signal for the
strange vector current form factors. The $\GMS$ results of the
Kentucky-Adelaide calculation were
fit to a dipole form form, and from that a value for $b_s^r$ was obtained.
There is no reason to
believe the dipole form is correct, as an insufficient number of low-$q^2$
values were computed to
constrain the behavior of the form factor in this region. At the same time,
the dispersion theory
values result from only the low-mass part of the spectral function, and
significant higher-mass
contributions are not ruled out.

With these caveats in mind, it is interesting to note that the results of
the Kentucky-Adelaide
lattice calculation of $\mu_s$ are consistent with the low-mass dispersion
relation value, but
rather substantial disagreement occurs for the magnetic and electric radii.
Even the
signs of the two
calculations of $<r^2_s>_{M,E}$ differ.  Clearly, there is room
for improvement in the theory. On the experimental side, completion of the
entire program of
PVES strange quark measurements is the only way to ensure that defensible
theoretical progress is
made.

\begin{table}
\begin{tabular}{|c|c|c|c|}
\hline
Method & $b_s+0.6b_8$ & $b_s^r$ & $c_s^r$
\\ \hline
Lattice/Dipole Fit & $-0.6\pm 0.1$  & $0.23\pm 0.13$ & $-0.06\pm 0.4$ \\ \hline
Low-mass Dispersion Relation & $-0.6$  & $-1.1$ & $3.4$ \\
\hline
\end{tabular}
\caption{Low-energy constants for strange magnetic moment
($b_s$),
strange magnetic radius ($b_s^r$), and strange electric radius ($c_s^r$).
Lattice QCD values are
obtained from Ref. ~\protect\cite{liu1}. \lq\lq Low-mass" dispersion
relation
contributions
obtained from Ref. \protect\cite{hammer2}.}
\label{tab1}
\end{table}

%\begin{table}
%\begin{center}
%\begin{tabular}{|c|c|c|c|}\hline
% Method & $b_s+0.6b_8$ & $b_s^r$ & $c_s^r$
%\\ \hline
%Lattice/Dipole Fit & $-0.6\pm 0.1$  & $0.23\pm 0.13$ & $-0.04\pm 0.4$ \\
%\hline
%Low-mass Dispersion Relation & $-0.6$  & $-1.1$ & $3.4$ \\ \hline
%\end{tabular}
%\end{center}
%\vspace{0.5cm}
%\caption{Low-energy constants for strange magnetic moment
%($b_s$),
%strange magnetic radius ($b_s^r$), and strange electric radius ($c_s$).
%Lattice QCD values are
%obtained from Ref.~\protect\cite{liu}. \lq\lq Low-mass" dispersion relation
%contributions
%obtained from Ref. \protect\cite{hammer}.}
%\label{tab1}
%\end{table}

\section{New Wrinkles: Hadron Structure}

While the current program of PVES measurements is far from complete, some
intriguing puzzles
have emerged from the results already reported, particularly in relation to
electroweak radiative
corrections. Theoretically, the challenge is to properly sort out the
interplay between
electroweak radiative corrections and strong interactions. In this context,
a new idea for a future
measurement of PV pion photoproduction on the $\Delta$ resonance has
emerged recently -- a
measurement which may shed new light on an old problem in hadronic weak
interactions.

\subsection{Electroweak radiative corrections}

It is now well known that the importance of electroweak radiative
corrections to the
axial vector form factor are enhanced relative to the naively expected
${\cal O}(\alpha/\pi)$
scale. This enhancement occurs because (a) the tree-level coupling is
suppressed by the small
vector coupling of the electron to the $Z^0$, $g_V^e=-1+4\sstw\approx
-0.1$, and (b) the
corrections themselves contain large logarithms. These logarithms only
appear in amplitudes
involving $\gamma$-exchange -- so they do not affect neutrino probes of the
axial vector hadronic
current. The large $\gamma$-exchange contributions can also be affected by
parity-violating
quark-quark interactions, which induce a parity-violating $\gamma NN$ coupling.
People sometimes refer to this coupling as the nucleon
anapole moment
coupling. The anapole moment {\em per se} is not a physical observable,
since its value
depends on the choice of electroweak gauge parameter, $\xi$. Moreover,
anapole moment effects can
never be distinguished from other electroweak radiative corrections in any
experiment. However, we
might still use the \lq\lq anapole moment" to refer to the
$\xi$-independent contribution of the
weak quark-quark interaction to the leading PV $\gamma NN$ coupling (for a
discussion see Ref. \cite{musolf1}).

It has been known for some
time that because the anapole moment (AM) effects involve hadronic weak
interactions, their
magnitude is uncertain\cite{musolf2}. Thus, the theoretical prediction for
$G_A^e$ -- the axial vector
form factor measured in PVES -- has a fairly sizeable error bar. Were one
to try and determine $\GMS$
from PV $ep$ scattering alone, the theoretical $G_A^e$ uncertainty would
add to the error bar for
$\GMS$. Hence, the SAMPLE collaboration has measured both the PV elastic
$ep$ and PV quasielastic
$eD$ asymmetries, which depend on different linear combinations of the two
form factors in
question and allow an experimental separation of $G_A^e(I=1)$ from $\GMS$.
The theoretical
uncertainty associated with $G_A^e(I=0)$ has been estimated to be
considerably smaller than for
the isovector component, so the isoscalar contribution will be neglected in
this
discussion\footnote{Of course, the estimate of uncertainty for $G_A^e(I=0)$
could be too small,
a possibility that should be investigated.}.

The published SAMPLE results\cite{sample1} indicate that the radiative
corrections to $G_A^e(I=1)$
are, indeed, large and have the same sign as given by the calculations of
Refs. \cite{musolf2,zhu3}.
However, the magnitude may be considerably larger than predicted. If this
result is confirmed by
future measurements, such as the lower-$Q^2$ SAMPLE deuterium experiment or
G0, one would like to
understand what is behind the enhancement. Several ideas have been mentioned:

\medskip

\noindent{\em Nuclear PV Effects}. In the extraction of $G_A^e(I=1)$ from
the deuterium asymmetry,
possible contributions to the PV $\gamma$-{\em nucleus} coupling from PV
$NN$ interactions have
not previously been taken into account. In contrast to PV effects in
elastic scattering, those
entering inelastic scattering can give rise to a $q^2$-independent
contribution to the asymmetry:
\begin{eqnarray}
A_{LR}({\rm elastic}) & = & A_1 q^2 +\cdots \\
A_{LR}({\rm inelastic}) &  &  A_0 + A_1 q^2 +\cdots \ \ \ .
\end{eqnarray}
The $A_1 q^2 +\cdots$ terms arise from the $Z^0$-exchange amplitude and
contain the usual electroweak
form factors. The $A_0$ term in $A_{LR}({\rm inelastic})$ is generated by
$\gamma$-exchange, where
the $\gamma$ couples to the nucleus through a PV transverse electric dipole
matrix element. The
presence of this term follows directly from an extension of Siegert's
theorem\cite{friar}, which
implies that for elastic scattering, all  transverse electric dipole matrix
elements are proportional
to $q^2$, while those entering inelastic transitions need not vanish at the
photon point.
In the case of $A_0$, PV nucleon-nucleon interactions are responsible for
the effect. People
(myself included) have speculated that for sufficiently small $|q^2|$, the
$A_0$ term may become quite
important relative to the $A_1 q^2$ term that contains $G_A^e$, and that the
larger-than-expected radiative corrections to $G_A^e$ may actually be a
result of neglecting the
$A_0$ contribution.

To address this question, two groups recently computed the
$q^2$-independent contribution
generated by nuclear PV \cite{paris,liu3}. The results of the two
calculations are consistent with each other and
indicate that the effect of nuclear PV is far too small to modify the value
of $G_A^e$ extracted from
the asymmetry. Generally speaking, one has\cite{liu2}
\begin{equation}
\left|{A_0\over A_1 q^2}\right| \sim 10^{-4}\ {m_N^2\over |q^2|} \ \ \ .
\end{equation}
At $|q^2|=0.1$ (GeV$/c)^2$, the ratio is about $10^{-3}$. Thus, if the
radiative corrections are truly
enhanced, then some other mechanism would have to be the culprit.

\medskip

\begin{figure}
\begin{center}
\resizebox{5. cm}{!}{\includegraphics*[200,360][420,730]{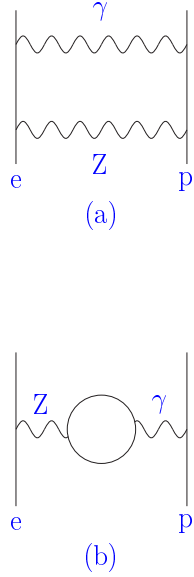}}
\caption{\label{Fig2}(a) $\gamma Z$ box graph contribution to
parity-violating $ep$ scattering.
(b) $Z\gamma$ mixing tensor contribution to parity-violating $ep$ scattering.}
\end{center}
\end{figure}

\noindent{\em Box Graphs}. One of the more sizeable contributions to the
axial vector radiative
corrections arises from the $Z$-$\gamma$ box diagrams (Fig. \ref{Fig2}a). The
size of this amplitude is
\begin{equation}
\label{eq:zgammabox}
{\cal M}_{\gamma Z\ {\rm Box}} = V_e\times \left[ a_0 A_N^{I=0} +
a_1 A_N^{I=1}\right] \ \   {G_F\over 2\sqrt{2}}
{\alpha\over
4\pi}
\left[\ln\frac{M_Z}{\Lambda} + C_{\gamma Z}^A(\Lambda)\right]\ \ \ ,
\end{equation}
where $V_e^\mu = {\bar e}\gamma^\mu e$, $A_{N\mu}^{I=1}={\bar
N}\gamma_\mu\gamma_5\tau_3 N$, and $A_{N\mu}^{I=0}={\bar
N}\gamma_\mu\gamma_5 N$.
The constants $a_i$ are given by $a_0\approx\frac{1}{6}(3F-D)(9-20\sstw)$
and $a_1=\frac{1}{2}(D+F)(1-4\sstw)$.
Note also the presence
of the large logarithm. Here, $\Lambda$ is a low-momentum cut-off, usually
taken to be
about one GeV,
corresponding to the transition between the perturbative and
non-perturbative domains. The
constant $C_{\gamma Z}^A(\Lambda)$ encodes all the physics of the loop
integral associated with
momenta $k<\Lambda$. This physics includes hadronic form factors at the
hadronic vertices as well
as the full tower of hadronic intermediate states in the hadron propagator.
To date, no one has
attempted to carry out a systematic treatment of all of these effects. Only
the nucleon
intermediate state -- along with the relevant form factors -- has been
included in model
computations\cite{marciano1,musolf2}. One might wonder then, whether
inclusion of other states, such
as the uncorrelated
$\pi N$, $\Delta$, $N^*$, etc, might lead to an enhancement of this
contribution. The problem is
an open one and remains to be tackled.

I like to emphasize that understanding this particular contribution could
have implications for
the interpretation for other precision electroweak measurements. For
example, in neutron
$\beta$-decay, the $W\gamma$ box graph produces a similar structure to that
of Eq.
(\ref{eq:zgammabox}). In this case, the $A({\rm lepton})\times V(N)$
amplitude is used to extract the
value of $V_{ud}$ and test the unitarity of the CKM matrix. A value for the
constant
$C_{\gamma W}^V(\Lambda)$ has been estimated, but as in the case of
$C_{\gamma Z}^A(\Lambda)$,
contributions from the full hadronic Green's function have not been
included. Were
$C_{\gamma Z}^A(\Lambda)$ large enough to give the enhanced radiative
corrections suggested by
SAMPLE, and were $C_{\gamma W}^V(\Lambda)$ to have a similar magnitude, one
would need to change the
value of $V_{ud}$ substantially and engender serious problems with CKM
unitarity. Although such a
scenario seems unlikely, one might nevertheless gain new insights into the
theoretical uncertainty
associated with $C_{\gamma W}^V(\Lambda)$ through a study of $C_{\gamma
W}^A(\Lambda)$.

>From a somewhat different perspective, the SAMPLE Collaboration has also
measured the
parity-conserving asymmetry associated with transversely polarized
electrons\cite{sample2}. This
asymmetry arises from the $2\gamma$-exchange box graph, which is also
sensitive to the low-energy part
of the nucleon Green's function. The experimental result for the asymmetry
does not agree with
theoretical predictions. Achieving a better understanding of the $Z\gamma$
box graph may shed new
light on this problem, and vice versa.

\medskip

\noindent{\em $Z\gamma$ Mixing}. The largest contribution to the axial
vector radiative corrections
comes from the $Z\gamma$ mixing tensor, $\Pi_{\gamma Z}$ (Fig.
\ref{Fig2}b). The
light quark
contribution to $\Pi_{\gamma Z}$ cannot be computed in perturbation theory,
since the quarks
have non-perturbative strong interactions. Traditionally, this problem is
solved by relating
the light quark component of $\Pi_{\gamma Z}$ to $\sigma(e^+ e^-\to{\rm
hadrons})$ using
a dispersion relation and SU(3) arguments. For purely leptonic scattering,
this approach ought to
suffice. For $eq$ scattering, however, one might worry about strong
interactions between the
quarks in $\Pi_{\gamma Z}$ and quarks in the proton. To some extent, these
interactions have
already been taken into account in anapole moment computations, but the
matching of the two
calculations onto each other remains to be worked out carefully. Again, it
seems unlikely that
some new effect might emerge when this matching has been achieved, but one
should think about it
nonetheless.

\subsection{PV $\Delta$ Electro- and Photoproduction: QCD Symmetries and
Weak Interactions}

The G0 Collaboration plans a study of the $q^2$-dependence of the axial
vector $N\to\Delta$
transition form factor, $G_A^{e\Delta}$. As in the case of $G_A^e$, the
transition axial vector
form factor will be modified by electroweak radiative corrections. The same
issues which apply
to $G_A^e$ will also apply to $G_A^{e\Delta}$ (for a detailed analysis, see
Ref. \cite{zhu4}). In
principle, these corrections could contain some $q^2$-dependence not
associated with the \lq\lq
primordial" form factor (as one might more directly measure with neutrino
scattering). A more serious
issue, however, arises because one is studying an inelastic process. For
the same reasons as
discussed above, the PV
$N\to\Delta$ asymmetry will receive a $q^2$-independent term as a
consequence of Siegert's theorem.
In this case, the relevant $E1$ matrix element is parameterized by an LEC
called
$d_\Delta$. At $q^2=0$ one
has\cite{zhu5}
\begin{equation}
A_{LR}^{N\to\Delta}(q^2=0) = -2{d_\Delta\over C_3^V}{m_N\over
\Lambda_\chi}\ \ \ ,
\end{equation}
where $C_3^V$ is the leading vector current transition form factor. The
presence of this term in
the PVES asymmetry could complicate the determination of the
$q^2$-dependence of $G_A^{e\Delta}$,
so it would be desirable to know it in some other way. In principle, a
study of PV photoproduction
could allow one to measure $d_\Delta$, as discussed by Jeff Martin
elsewhere in these proceedings.

A determination of $d_\Delta$ would also be interesting from another
standpoint -- namely, an old
problem in nonleptonic weak interactions. In the $\Delta S=1$ sector, one
has considerable
knowledge of hyperon decays. The mesonic decays depend on both S- and
P-wave amplitudes. To date,
it has not been possible to obtain a successful, simultaneous description
of both amplitudes in
ChPT. Another problem appears with the $\Delta S=1$ radiative decays: $B\to
B'\gamma$. Here, one
measures PV asymmetries $\alpha_{BB'}$ associated with the direction of the
photon relative to the
spin of the initially polarized hyperon. These asymmetries arise from the
interference of an E1 and
an M1 amplitude. In the limit of exact SU(3) flavor symmetry, the octet of
baryons is degenerate and
the E1 amplitude vanishes -- a result known as Hara's theorem\cite{hara}.
In the real world, SU(3)
symmetry is broken by the $m_s$ vs. $m_u+m_d$ mass difference. Thus, one
would expect the asymmetries
to have a size $\alpha_{BB'}\sim m_s/\Lambda_\chi\sim 0.15$. The
experimentally measured
$\alpha_{BB'}$ are  factors of four to five larger than this naive
expectation. Based on general
symmetry considerations, one has no way of explaining what is responsible
for these enhancements.

It would appear, then, that the application of QCD-based symmetries, such
as chiral symmetry and
SU(3) symmetry, to the theoretical analysis of $\Delta S=1$ non-leptonic
weak interactions fails.
One would like to know if this failure is a consequence of having strange
quarks involved in
the decay processes, or if it is a signature of a more fundamental feature
of the interplay of
strong and weak interactions.

To that end, a study of $d_\Delta$ could provide new insight. The
transition induced by the
$d_\Delta$ operator is the $\Delta S=0$ analog of the $\Delta S=1$ E1
amplitudes governing the
$\alpha_{BB'}$. If the enhancements of the latter are not due to strange
quarks but a more general
feature of nonleptonic weak interactions, then one ought to observe an
enhanced asymmetry for
PV $\Delta$ photoproduction, which involves no strange quarks. On the other
hand, if strangeness
is the key to the breakdown of symmetry considerations, then the $\Delta
S=0$ asymmetry ought to
have its natural size.

\begin{figure}
%\hspace{0.05in}
\begin{center}
\resizebox{5. cm}{!}{\includegraphics*[200,360][420,730]{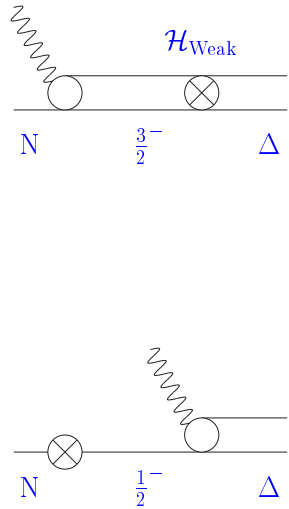}}
\caption{\label{Fig3}Resonance mixing contributions to parity-violating
$\Delta$ photoproduction.}
\end{center}
\end{figure}

Recently, Holstein and Borasoy have developed a model that allows us to
make this idea
quantitative\cite{holstein1}. The idea is that the $\Delta S=1$ hyperon
decays are affected by mixing
of excited baryons into the ground states due to the hadronic weak
interaction . In
order to resolve the S-wave/P-wave problem, the mixing matrix elements
$W_{BB'}(\Delta S=1)$ must have a
characteristic size larger than one might have naively expected. After
determining what the
$W_{BB'}(\Delta S=1)$ must be to fix the S-wave/P-wave inconstency,
Holstein and
Borasoy used them to predict the
$\alpha_{BB'}$ via resonance-mixing processes. Interestingly, the
asymmetries are
significantly enhanced by this mechanism, thereby helping to close the gap
between the naive
symmetry expectation and the experimental results.

In principle, the same kind of weak interaction mixing could occur in the
$\Delta S=0$ sector (see
Fig. \ref{Fig3}). If the mixing matrix elements $W_{BB'}(\Delta S=0)$ have
the same characteristic
size as for the
$\Delta S=1$ case, then one would expect $d_\Delta$ to be significantly
enhanced -- by factors of
25 or more -- compared to one's naive expectation. We already know from
studies of PV $pp$
scattering that PV
$N\to NM$ amplitudes
($M$ is a meson) are of the same size as the $\Delta S=1$ $B\to B'\pi$
amplitudes, so it is
not unreasonable to expect the $W_{BB'}$ to have the same scale in the
$\Delta S=0$ and
$\Delta S=1$ sectors. In this case, $A_{LR}^{N\to\Delta}(q^2=0)$ would be
rather large -- on the
order of ppm -- and one could measure it with a reasonable amount of beam
time. On the other
hand, if one looked for an asymmetry of this size and didn't see it, one
might raise doubts about
the Holstein/Borasoy proposal and, more generally, conclude that the
problem with symmetry truly
does hinge on having a strange quark involved.

\section{New Wrinkles: The Standard Model and Beyond}

The first few PVES measurements were intended to test the Standard Model
(SM). Since then, the SM
has been so well-tested in a variety of ways that we can, with a high
degree of confidence, use
the weak $eq$ interaction as a well-understood probe of hadron and nuclear
structure. The recent advances
in this field -- both experimental and theoretical -- have now afforded us
an opportunity to come
full circle and again study fundamental aspects of the weak interaction.

Of course, there is little doubt that the SM is a successful model
describing weak interactions far
below the Planck scale. However, there exist many conceptual reasons for
believing that the SM
must be embedded in a larger theory. The quest to discover this \lq\lq new"
Standard Model is at
the forefront of both particle physics and nuclear physics. Unravelling the
nature of the neutrino
is clearly direction in which nuclear and particle physicists are pursuing
this goal. Future
collider experiments at the Large Hadron Collider and the Linear Collider
will seek direct
evidence for new physics. A third, and no less interesting, avenue is to
carry out highly precise
measurements of electroweak observables. Some such observables -- such as
the permanent electric
dipole moments of the electron, neutron, and neutral atoms -- are so small
in the SM that
the measurement of a nonzero effect would provide smoking gun evidence for
new physics. Others,
such as the PV asymmetries of interest here, have non-zero values in the SM, so
one must attempt to
probe for tiny deviations from the SM predictions. The presence or absence
of such deviations --
when taken in conjuction with other precision measurements -- can provide
important clues about
the structure of the new SM.

To illustrate, recall how fits to precision electroweak observables
predicted a range for the mass of the top quark prior to its discovery at
the Tevatron. The
subsequent discovery of a top quark with mass lying in this range provided
beautiful confirmation
of the SM at the level of quantum corrections. The same kind of synergy
between indirect searches
in precision electroweak measurements and direct collider searches will
undoubtedly provide new
insights about whatever lies beyond the SM.

>From this standpoint, the quantity of interest to PVES is $Q_W^f$, the weak
charge of a fermion or
system of fermions. At tree level in the SM, it is the coupling of the
$Z$-boson to the fermion
$f$. It also governs the strength of the effective $A(e)\times V(f)$
interaction at $Q^2=0$:
\begin{equation}
{\cal L}^{\rm eff}_{PV} = -{G_F\over 2\sqrt{2}} Q_W^f{\bar
e}\gamma^\mu\gamma_5 e {\bar f}\gamma_\mu
f\
\ \ ,
\end{equation}
where, at tree-level in the SM, one has
\begin{equation}
\label{eq:qweak}
Q_W^f = 2I_3^f -4 Q_f\sstw\ \ \ .
\end{equation}
For both the electron and the proton, the magnitude of the weak charge is
suppressed:
$Q_W^p=-Q_W^e\approx 0.1$. Because of this suppression, $Q_W^{p,e}$ are
rather transparent to
new physics. As we have heard elsewhere in this conference, the goals of
the PV M\" oller
experiment at SLAC and the PV $ep$ experiment at Jefferson Lab are to
measure these two weak
charges with some hope of finding evidence for new physics.

We have also heard that even within the SM, these two measurements are
quite interesting.
The value of the effective weak mixing angle appearing in Eq.
(\ref{eq:qweak}) actually depends
on the scale at which one is carrying out the measurement. The SM predicts
this scale-dependence.
The SM prediction at the Z-pole has been confirmed with high precision, but
there exist few
precise determinations of $\sstw(q^2)$ below the $Z^0$-pole. The cesium
atomic PV experiment carried
out by the Boulder group\cite{wieman} probed the weak mixing angle -- via
the cesium nucleus weak
charge ($Q_W^{\rm Cs}$) -- at a very low scale. Although early indications
suggested a substantial
deviation from the SM, recent developments in atomic theory now imply the
cesium atomic PV result
is consistent with the SM. Since the extraction of $Q_W^{\rm Cs}$ from the
measured PV transition
rate depends on atomic structure calculations, there has always been some
question about the
theoretical uncertainty associated with this extraction. In the past couple
of years, atomic
theorists have computed several sub-one percent corrections -- such as the
Breit interaction
correction\cite{derevianko}, Uehling potential\cite{johnson}, and
nucleus-enhanced QED radiative
corrections\cite{flambaum,sushkov}. As of this writing, there appears to be
an emerging consensus that
all of the important corrections have been properly taken into account and
that the value of $Q_W^{\rm
Cs}$ should be stable at the level of the experimental precision (about 0.4\%).

A value for $\sstw(q^2)$ at somewhat higher scales has been obtained by the
NuTeV Collaboration
using deep inelastic neutrino-nucleus scattering\cite{nutev}. The result
indicates a roughly 3$\sigma$
deviation from the value predicted in the SM. As in the case of cesium APV,
however, there exists
some debate about the theoretical uncertainties associated with this
extraction. It has been argued,
for example, that small changes in parton distribution
functions due to nuclear shadowing
and isospin-breaking could substantially change the SM prediction for the
neutrino scattering cross
sections and reduce the significance of the NuTeV
anomaly\cite{miller}. The Collaboration has
responded to this suggestion with a vigorous rebuttal\cite{zeller}. When
the dust finally settles, one
will have either learned something about parton distributions in nuclei or
have striking evidence for
new physics.

In the context of these other measurements, the SLAC and JLab measurements
offer several
advantages. First, they are being undertaken at values of $q^2$ lying
between the cesium atomic PV
and NuTeV scales. Second, they are both being performed at roughly the {\em
same} scale:
$Q^2\approx 0.03$ (GeV$/c)^2$. Third, they are theoretically clean, so
their intepretation will
not be subject to the kinds of controversies that have affected cesium
atomic PV and the NuTeV
result. This point is particularly important, so I will spend some time on
it below. Finally, the
two measurements are complementary. One is purely leptonic, while the other
is semileptonic. As I
will also discuss, a comparsion of the two measurements can provide
an interesting \lq\lq
diagnostic" tool for probing new physics.

\subsection{Theoretical Interpretability}

In principle, the PV M\" oller experiment, being purely leptonic, is the
theoretically cleanest
observable. There is some theoretical uncertainty in the SM prediction
arising from the presence
of light quark loops in  $\Pi_{\gamma Z}$\footnote{The issue raise earlier
about interactions between
these quarks and target quarks is less severe here. At $q^2=0$, these
effects are constrained by
current conservation.}. This uncertainty is common to both
$Q_W^e$ and $Q_W^p$. Traditionally, the light quark contribution to
$\Pi_{\gamma Z}$ is obtained by relating it to $\sigma(e^+ e^-\to{\rm
hadrons})$ via a dispersion
relation. In order to do so, one must also invoke SU(3) flavor arguments.
The uncertainty associated
with this procedure was estimated some time ago by Marciano and
Sirlin\cite{marciano1}, and it falls
below the expected experimental precision for both weak charge measurements.

A second source of theoretical uncertainty arises from the extraction of
$Q_W^{e,p}$ from the
measured asymmetries. In the case of the M\" oller experiment, one must be
able to reliably subtract
out contributions from the elastic and inelastic $ep$ asymmetries. Although
the number of $ep$
events is small when compared with the number of M\" oller events, the
inelastic $ep$ asymmetries
can be relatively large. The basic problem is that we do not know in detail
what the inelastic
asymmetries are. One strategy for circumventing it is to measure the total
asymmetry in a kinematic
region where the inelastic $ep$ asymmetry is dominant, and then extrapolate
to the
kinematics of the M\" oller
measurement. To the extent that various components of the inelastic
asymmetry evolve with
the relevant kinematic variables in a well
understood way, this strategy should reduce the corresponding $ep$
uncertainty to an acceptable level.

In the case of $Q_W^p$, the design of the detector should eliminate any
sizeable contribution from
inelastic events. However, the elastic $ep$ asymmetry itself receives
hadronic contributions in the
guise of nucleon form factors. The associated uncertainty can also be
reduced to an acceptable level
by drawing on other measurements. Specifically, the asymmetry has the form
\begin{equation}
A_{LR}^{ep} = {G_F Q^2\over 4\sqrt{2}\pi\alpha}\left[Q_W^p +
F^p(Q^2,\theta)\right]\ \ \ ,
\end{equation}
where $F^p(Q^2,\theta)$ is a function of nucleon form factors, both
electromagnetic and strange.
At forward angles and low-$Q^2$, $F^p\sim Q^2$, yielding an overall $Q^4$
contribution to the
asymmetry. In contrast the $Q_W^p$ term goes as $Q^2$. The strategy, then,
is to use the existing
program of PV measurements to constrain the $Q^2$-dependence of $F^p$ and
extrapolate to very
low-$Q^2$ where the Q-Weak measurement will be performed and where the
$Q_W^p$ term gives the
dominant contribution to the asymmetry. This strategy can be implemented in
a rigorous way by using
ChPT to express all the form factors in $F^p$ in terms of their known
quark-mass dependence plus
LEC's. The existing program of experiments, plus the world data set for the
electromagnetic form
factors, will determine the LEC's with sufficient precision to make the
uncertainty in the low-$Q^2$
asymmetry sufficiently small. No additional theoretical input on nucleon
structure will be needed.

Potentially more problematic considerations arise in arriving at the SM
prediction for $Q_W^p$
itself. In particular, the two boson-exchange box graphs depend on hadronic
physics via the nucleon
Green's function in the box. Fortunately, we are able to constrain the
theoretical uncertainty
associated with these effects to be acceptably small.

The largest box graph contribution involves the exchange of two $W$-bosons.
The total correction is
about 26\% of the tree-level value, largely due to the absence of the
$1-4\sstw$ factor from the
${\cal M}_{WW}$, the WW box amplitude. The latter is dominated by
intermediate states of momentum
$k\sim M_W$, so the QCD corrections are perturbative. The latter can be
computed using
a combination of current algebra and the operator product expansion. The
result is \cite{erler}
\begin{equation}
{\cal M}_{WW} =  -A(e)\times V(p) \ \ {G_F\over 2\sqrt{2}}
{{\hat\alpha}\over 4\pi {\hat s}^2}\left[2+
5\left(1-{\alpha_s(M_W)\over\pi}\right)\right]\ \ \ ,
\end{equation}
where ${\hat\alpha}$ is the fine structure constant evaluated at a scale
$\mu=M_W$ in the
${\overline{MS}}$ scheme and ${\hat s}^2$ is the corresponding definition
of $\sstw$. Inclusion of the
${\cal O}(\alpha_s)$ contributions generate a $\sim -0.7 \%$ correction to
$Q_W^p$, and higher-order
corrections in $\alpha_s$ will be at roughly an order of magnitude less
important. Note that the QCD
corrections involve no large logarithms, since the currents involved are
conserved (or partially
conserved). A similar correction applies to the ZZ box graphs, but the
effect in this case is much
smaller -- roughly $-0.1\%$ in $Q_W^p$.

In contrast, the $\gamma Z$ box graphs receive contributions from both low-
and high-momentum scales.
This sensitivity to the different scales is reflected in the presence of a
large logarithm in the
amplitude:
\begin{equation}
{\cal M}_{\gamma Z} = -A(e)\times V(p) \ \ {G_F\over 2\sqrt{2}}
{5{\hat\alpha}\over 2\pi} (1-4{\hat
s}^2)\left[\ln\left({M_Z^2\over\Lambda^2}\right) + C_{\gamma
Z}^V(\Lambda)\right]\ \ \ ,
\end{equation}
where I have omitted contributions suppressed by powers of $p_e/M_Z$. As in
Eq. (\ref{eq:zgammabox}),
$\Lambda$ is a scale associated with the transition from the
non-perturbative to the perturbative
region and is on the order of 1 GeV. At present,
we cannot reliably compute contributions to the loop
integral arising from momenta below this scale. Instead, we parameterize
their effects by the constant
$C_{\gamma Z}^V(\Lambda)$. Fortunately, the effect of this unknown term is
suppressed by the overall
$1-4{\hat s}^2$ prefactor, so we can live with a fairly large uncertainty
in $C_{\gamma Z}$ without
affecting the interpretation of $Q_W^p$. The presence of this prefactor is,
itself, something of a
fortunate accident. It arises from the sum of the box and crossed-box
graphs. The spacetime structure
of this sum dictates that the $A(e)\times V(p)$ amplitude arises from the
antisymmetric product of the
electron electromagnetic current and the vector part of the electron weak
neutral current. The latter
contains the $1-4{\hat s}^2$ vector coupling of the electron to the $Z^0$,
thereby producing the
suppression factor in ${\cal M}_{\gamma Z}$.

To estimate the uncertainty associated with $C_{\gamma Z}^V$ we can follow
a philosophy adopted by
Sirlin in his classic analysis of neutron $\beta$-decay\cite{sirlin}. In
that case, a similar unknown
constant arises from the $\gamma W$ box graph. The size and uncertainty of
$C_{\gamma W}^V$ is
constrained because the vector part of the decay amplitude is used to determine
$V_{ud}$\footnote{This amplitude is actually known most precisely from
superallowed Fermi nuclear
$\beta$-decay, though new experiments using ultracold neutrons should
provide competitive
determinations.}. In order for
$V_{ud}$ to be consistent with the unitarity of the CKM matrix, the
uncertainty from $C_{\gamma W}^V$
cannot be  larger than about $\pm 2$. The hadronic dynamics responsible for
$C_{\gamma W}^V$ and
$C_{\gamma Z}^V$ are similar, so one would expect the uncertainty in the
two to be comparable. The
corresponding uncertainty in $Q_W^p$ is about $\pm 1.5\%$. A siginficantly
larger uncertainty in
$C_{\gamma W}^V$ and $C_{\gamma Z}^V$ would imply unacceptably large
deviations from CKM unitarity.
Of course, one should pursue computations of these quantities from first
principles -- possibly using
lattice QCD. However, the fortuitous presence of the $1-4{\hat s}^2$
suppression factor has given us
a sizeable zone of theoretical safety.

One other hadronic effect which one might worry about for $Q_W^p$ is the
impact of isospin admixtures
into the proton wavefunction. On general grounds, one would expect
isospin-breaking effects to be of
${\cal O}(\alpha)$ and, therefore, a source of concern. At $Q^2=0$,
however, a diagonal version of the
Ademollo-Gatto theorem\cite{ademollo} implies that isospin-breaking effects
in the matrix element of
the hadronic vector current vanish identically\cite{erler}. They only turn
on for non-zero $q^2$. In
this case, they will be effectively contained in the measured form factor
term $F^p$ and constrained
-- along with all other
$q^2$-dependent effects -- by experiment.

\subsection{New Physics}

Given the theoretical interpretability of the two weak charge measurements,
one can credibly view
them as probes physics beyond the SM. I have been thinking quite a bit
about supersymmetry (SUSY)
recently, so I will focus my discussion on this brand of new physics. SUSY
remains one of the most
strongly-motivated extensions of the Standard Model for a number of
reasons: it gives a solution
to the \lq\lq hierarchy problem" in particle physics (essentially, the question
as to why the electroweak
scale is stable); it generates unification of the electroweak and strong
couplings close to the
Planck scale; it gives a natural candidate for cold dark matter (the
neutralino, ${\tilde\chi}^0$);
and it contains new sources of CP violation that could help us account for
the observed predominance
of matter over anti-matter (the baryon asymmetry of the universe). Of
course, since no one has yet
seen any of the \lq\lq superpartners" of SM particles, they must be heavier
than the particles in the
SM. Consequently, SUSY must be a broken symmetry. The breaking cannot be
too large, however;
otherwise, the hierarchy problem re-emerges.

Future collider experiments at the LHC and LC will go looking for the heavy
superpartners. One might
also ask about observing their effects indirectly. For example,
superpartners can contribute to
precision electroweak observables through radiative corrections. Perhaps,
the most well-known example
of such effects are superpartner loop contributions to the muon anomalous
magnetic moment. On general
grounds, one expects the size of SUSY radiative corrections to go as
$\alpha/\pi(M/{\tilde M})^2$,
where $M$ is a SM mass and ${\tilde M}$ is a superpartner mass. For the
muon $g-2$, one has $M=m_\mu$.
Taking ${\tilde M}\sim 100$ GeV, one gets an effect of the order of
$10^{-9}$ -- roughly the size of
the current experimental error bar. The possible significant deviation of
$(g-2)_\mu$ from the SM
prediction would be consistent with SUSY loop contributions containing
superpartners of mass of a few
hundred GeV. For this reason, the particle physics community has become
quite excited by new results
for the muon anomaly\footnote{There remains some controversy over the
theoretical uncertainty in the
SM prediction associated with hadronic contributions. I will not discuss
this controversy here.}.

For weak interaction processes, such as $\beta$-decay and PVES, one has
$M=M_W$. Consequently, the
magnitude of SUSY loop effects, {\em relative} to the SM prediction, is
generically of order
$10^{-3}$. However, since $Q_W^{e,p}$ are suppressed in the SM, one
actually needs a precision of
order a few percent -- rather than a few tenths of a percent -- to be
sensitive to SUSY radiative
corrections. For this reason, it is interesting to study the effect of
these radiative corrections on
the weak charges. To that end, we need to modify our formula for $Q_W^f$ as
follows:
\begin{equation}
Q_W^f = \rho_{PV}\left( 2 I_3^f - 4Q_f\kappa_{PV}{\hat s}^2\right)
+\lambda_f\ \ \ .
\end{equation}
Here, $\rho_{PV}$ and $\kappa_{PV}$ are universal, in that they do not
depend on the species of
fermion being probed with PVES. The quantity $\lambda_f$, in contrast, is
species-dependent. At
tree-level, one has $\rho_{PV} =1=\kappa_{PV}$ and $\lambda_f=0$. Inclusion
of loops lead to
deviations from these values. In particular, one can think of
$\delta\kappa_{PV}$ as the change in
the apparent weak mixing angle due to loop effects.  The effects of new
physics on $\kappa_{PV}$
arise in two ways: first, one has a change in the prediction for ${\hat
s}^2$ due to its defintion in
terms of other more precisely known quantities ($\alpha$, $G_\mu$, and
$M_Z$). This definition
gets modified by new physics. Second, $\delta\kappa_{PV}$ receives
contributions directly from the PV
$ef$ amplitudes.

In order to analyze the corrections $\delta\rho_{PV}^{\rm susy}$,
$\delta\kappa_{PV}^{\rm susy}$, and
$\lambda_f^{\rm susy}$ one needs to compute a large number of loop graphs.
Moreover, one has to take
into account the effects of SUSY-breaking. In general, this task is
non-trivial. In the Minimal
Supersymmetric Standard Model (MSSM), SUSY-breaking is governed by 105
parameters. To simplify the
problem -- and to build a real theory -- people have invented models of
SUSY-breaking, such as
supergravity, which reduce the number of independent parameters to a
handful. Whether or not these
model assumptions are entirely consistent with electroweak data is not
entirely clear. We recently
completed an analysis of charged current observables and found that the
superpartner spectrum implied
by SUSY-breaking models may not be consistent with the precision
data\cite{kurylov}. For
this reason, it is
desirable to carry out a model-independent analysis of SUSY radiative
corrections, thereby avoiding
the simplifying assumptions of SUSY-breaking models.

\begin{figure}
%\hspace{0.05in}
\begin{center}
\resizebox{11. cm}{!}{\includegraphics*[0,0][480,380]{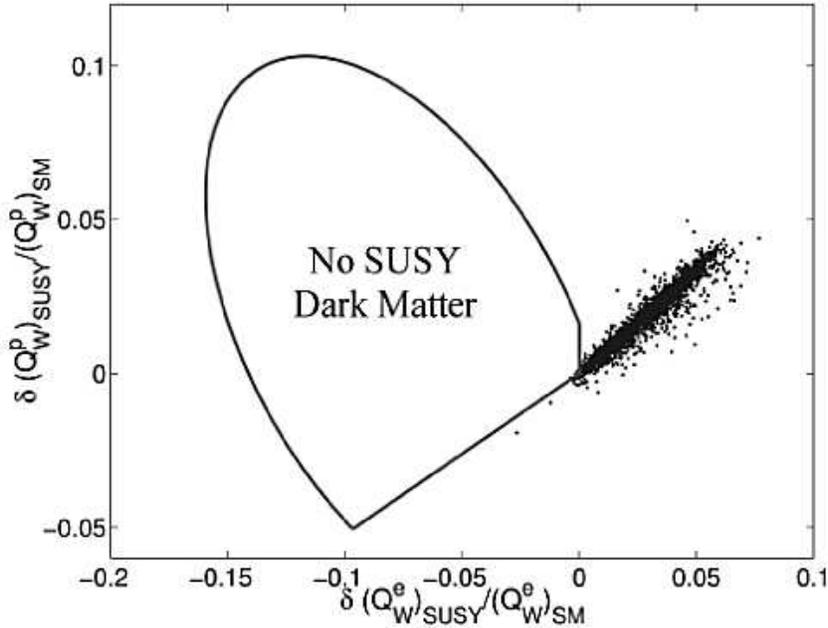}}
\caption{\label{Fig4}Relative shifts in the electron and proton weak
charges due to supersymmetric
effects. Dots indicate loop contributions, where each point corresponds to
a randomly-selected set of
soft SUSY-breaking parameters. Interior of truncated ellipse denotes
possible shifts from R
parity-violating interactions.}
\end{center}
\end{figure}

My collaborators and I have completed such a model-independent analysis of
SUSY radiative corrections
to $Q_W^{e,p}$ in the MSSM\cite{shufang}. We found that the effects on the
two weak charges are highly
correlated, and that the {\em relative} sign of the correction in both
cases is positive over nearly
all of the available SUSY-breaking parameter space. This correlation --
shown in Fig. \ref{Fig4} --
arises because the dominant effect occurs via $\delta\kappa_{PV}^{\rm
susy}$, with some scatter due
to the effects of
$\delta\rho_{PV}^{\rm susy}$ and $\lambda_f^{\rm susy}$. Moreover, the
potential
size of the SUSY
radiative corrections could be as large as the projected experimental error
bars. What this means is
that if both experiments come in consistent with the SM, then one could not
say much about the
SUSY-breaking parameters. However, deviations from the SM of the order of
$2\sigma$ or more
would -- depending on the relative sign of the effect -- start to either
favor or exclude parts the
parameter space. The situation is helped somewhat by the correlation
between the two weak charges in
SUSY. The two measurements together really act like one determination of
$\delta\kappa_{PV}^{\rm
susy}$ with better statistics than obtained with either alone.
Nevertheless, it would be advantageous
--looking further down the road -- to pursue even more precise measurements
of the weak charges.

The effects of SUSY can also arise at tree-level if one allows for the
violation of lepton
number (L) and/or baryon number (B). If B-L is conserved, then the MSSM has
another symmetry called
\lq\lq R parity". Conservation of R parity is not automatic; it has to be
imposed by hand. There are
some phenomenological constraints. The absence of proton decay highly
constrains the possibility of
B-violating interactions. However, the constraints on L-violating sources
of R-parity violation (RPV)
are much less severe. The existence of any such interactions would,
however, sound the death knell
for SUSY dark matter, since the lightest superpartner (presumably the
${\tilde\chi}^0$) would no
longer be stable. Thus, one would like to know from experiment whether or
not the MSSM
respects R parity.

To that end, a comparison of $Q_W^e$ and $Q_W^p$ measurements could be
quite interesting. In Fig. \ref{Fig4},
I show the possible deviations of these quantities from the SM values
due to RPV effects. The
interior of the truncated ellipse is the region allowed by other precision
measurements at 95\%
confidence. Note that for most of this region, the relative sign of the
deviation in $Q_W^e$ and
$Q_W^p$ is negative -- in contrast to the situation for SUSY loops. In
fact, there is nearly no
overlap between the regions in this plane associated with the different
effects. Thus, a comparison
of these two measurements could, in principle, serve as a \lq\lq
diagnostic" for SUSY with our
without R parity.

In fact, the diagnostic potential of these two experiments is even greater
if one considers other
models of new physics. For example, E$_6$ grand unified theories allow
for the possibility of a
relatively low mass, extra neutral gauge boson\footnote{The E$_6$ models
are actually a very broad
class of models, though they do not encompass all possibilities for a
\lq\lq low-mass" $Z'$.}. In
these models, the effect of a $Z'$ on $Q_W^e$ and $Q_W^e$ would also be
correlated (having the same
relative sign). However, a sizeable shift in these weak charges would also
imply a sizeable
deviation of the cesium weak charge from the SM prediction. Since $Q_W^{\rm
Cs}$ currently agrees
with the SM, this scenario appears to be disfavored. In contrast, the
effects of SUSY loops on
$Q_W^{\rm Cs}$ is tiny -- below the current experimental plus theoretical
error -- due to
cancellations between up- and down-quark contributions. Thus, one could
still have sizeable
loop-induced shifts in $Q_W^{e,p}$ without being inconsistent with cesium
atomic PV. Similarly, the
allowed RPV region shown in Fig. \ref{Fig4} already incorporates the
constraints from the $Q_W^{\rm
Cs}$ determination.

\section{Conclusions}

I am convinced that PVES will remain a  highly interesting field for some
time to come. In
addition to providing a definitive answer to the question about
strangeness, PVES has also opened up
possibilities for studying other aspects of hadronic and nuclear
structure\footnote{In the interest
of time, I have not discussed the important measurement of the neutron
distribution in lead \cite{michaels}.}. These studies rely on our present
understanding of the weak neutral
current interaction in order to probe novel aspects of low-energy QCD. At
the same time, the success
of this program has opened the way for new experiments to study the weak
interaction itself. Looking
further into the future, one might anticipate new PV deep inelastic
scattering experiments after the
CEBAF upgrade that may offer new glimpses of higher-twist physics, or even
a more precise
measurement of the M\" oller asymmetry\footnote{I am indebted to Paul
Souder, Paul Reimer, and Dave
Mack for a discussion of these possibilities.}. Suffice it to say, however,
that completion of the
current program of PVES measurements is essential to the long-term success
of the field.

\medskip

\noindent{\bf ACKNOWLEDGEMENTS}

I would like to thank the organizers of this workshop and the support staff
for their hospitality
during my stay in Mainz. I also thank A. Kurylov for assistance in preparing
this manuscript. This work was
supported in part under U.S. Department of Energy contracts
\# DE-FG03-02ER41215 and by the National Science Foundation under award
PHY00-71856.

\end{document}